\documentclass[referee]{raa}            

\usepackage{graphicx,times}             
\usepackage{natbib}
\bibpunct{(}{)}{;}{a}{}{,}

\usepackage[colorlinks = true, linkcolor = green, anchorcolor = red, citecolor = blue, filecolor = red, pagecolor = red, urlcolor = red]{hyperref}

\usepackage{upgreek}
\usepackage{soul}
\usepackage{CJK}

\begin{document}
\begin{CJK}{UTF8}{gbsn}
  \title{A catalog of new blue stragglers in open clusters with {\it Gaia} DR3}
  
   \volnopage{Vol.0 (20xx) No.0, 000--000}      
   \setcounter{page}{1}          

\author{Songmei Qin\inst{1,2,3},
          Jing Zhong \inst{1},
          Friedrich Anders \inst{3,4,5},
          Lola Balaguer-N\'u\~nez \inst{3,4,5},
          Chunyan Li\inst{6},
          Yueyue Jiang\inst{1,2,3},
          Guimei Liu\inst{7,2}，
          Tong Tang\inst{7,2}，
          Li Chen\inst{1,2}}

   \institute{Astrophysics Division, Shanghai Astronomical Observatory, Chinese Academy of Sciences, 80 Nandan Road, Shanghai 200030, PR China; {\it qinsongmei@shao.ac.cn,jzhong@shao.ac.cn,fanders@fqa.ub.edu,lbalaguer@fqa.ub.edu}\\
   \and
   School of Astronomy and Space Science, University of Chinese Academy of Sciences, No. 19A, Yuquan Road, Beijing 100049, PR China; \\
    \and 
    Institut de Ci\`encies del Cosmos, Universitat de Barcelona (ICCUB), Mart\'i i Franqu\`es 1, 08028 Barcelona, Spain;\\
    \and
    Departament de F\'isica Qu\`antica i Astrof\'isica (FQA), Universitat de Barcelona (UB), C Mart\'i i Franqu\`es, 1, 08028 Barcelona, Spain\\
    \and
    Institut d'Estudis Espacials de Catalunya (IEEC), Edifici RDIT, Campus UPC, 08860 Castelldefels (Barcelona), Spain\\
    \and
    Shanghai NanHui High School, No. 288 Xuehai Road, Shanghai 200120, PR China\\
    \and
    Xinjiang Astronomical Observatory, Chinese Academy of Sciences, Urumqi 830011, China\\
    \vs\no
   {\small Received~~20xx month day; accepted~~20xx~~month day}}


  \abstract
   {The high-precision {\it Gaia} data release 3 (DR3) enables the discovery of numerous open clusters in the Milky Way, providing an excellent opportunity to search for blue straggler stars in open clusters and investigate their formation and evolution in these environments. Using the member stars from literature open cluster catalogs, we visually inspected the color-magnitude diagram (CMD) of each cluster and selected cluster candidates that potentially host blue stragglers. We then reassessed cluster memberships using the {\tt pyUPMASK} algorithm with {\it Gaia} DR3 and performed isochrone fitting to derive physical parameters for each cluster, including age, distance modulus, mean reddening, and metallicity. Finally, we empirically identified straggler stars based on their positions relative to the best-fitting isochrone, zero-age main sequence (ZAMS), and equal-mass binary sequence on the CMD. In total, we identified 272 new straggler stars in 99 open clusters, comprising 153 blue stragglers, 98 probable blue stragglers, and 21 yellow stragglers. Compared to the reported blue straggler catalogs based on earlier {\it Gaia} data, our results increase the number of open clusters with stragglers in the Milky Way by 22.2\%, and the total number of blue stragglers by 11.2\%.
   \keywords{catalogs---stars: blue stragglers---Galaxy: open clusters and associations: general}
   }
   
      \authorrunning{Qin et al}            
   \titlerunning{Discovery of New Blue Stragglers}  

   \maketitle

\section{Introduction}\label{sec:intro}

\defcitealias{2021A&A...650A..67R}{R21}
\defcitealias{2021MNRAS.507.1699J}{J21}
\defcitealias{2023A&A...672A..81L}{L23}
\defcitealias{2023A&A...673A.114H}{HR23}

A mysterious class of stars in star clusters apparently {\it straggle} in their evolution, making them bluer and more luminous than the main-sequence turn-off stars in the color-magnitude diagram (CMD). These so-called blue straggler stars (BSSs) were initially detected in the CMD of the globular cluster M3~\citep{Sandage_1953}. Later, many studies have revealed that BSSs are commonly found in globular clusters~\citep{2004ApJ...604L.109P,2007ApJ...661..210L}, intermediate-age to old open clusters (OCs)~\citep{2021A&A...650A..67R,2021MNRAS.507.1699J,2023A&A...672A..81L}, and even field stars~\citep{2000AJ....120.1014P,2015ApJ...801..116S} of our Milky Way and dwarf galaxies~\citep{2007A&A...468..973M,2012MNRAS.421..960S}. BSSs have challenged the standard theories of stellar evolution, especially the effect of binary interaction. Meanwhile, BSSs within star clusters play an important role in investigating the dynamic evolution of their host clusters. 

A large number of studies have proposed various mechanisms for BSS origin~\citep{1964MNRAS.128..147M,1966ApJ...144..968I,1976ApL....17...87H,1993PASP..105.1081S}, and two prevailing scenarios have been widely accepted recently: 1) direct stellar collision and subsequent coalescence and 2) binary mass transfer and mergers. \citet{1976ApL....17...87H} firstly suggested BSSs may originate from stellar collisions between main sequence stars or between giants and main sequence stars.
Collisionally-induced BSSs undergo complete mixing and are expected to exhibit a nearly uniform helium distribution~\citep{1987ApJ...323..614B}. Moreover, the collision probability depends on stellar density, velocity dispersion, and binary fraction in clusters~\citep{1976ApL....17...87H,1999ApJ...513..428S}. Hence, collisionally-induced BSSs are more common in high-density environments such as GCs and the cores of the most massive OCs. The other possible scenario is mass transfer in a binary system~\citep{1964MNRAS.128..147M}, where a star accretes additional material (primarily hydrogen) from its companion. This process prolongs its main sequence lifetime, causing it become bluer and brighter than typical main sequence stars of the same age and mass, which ultimately produces a blue straggler in the binary system. Subsequently, a merger may occur, resulting in the formation of a BSS as a single star~\citep{1990AJ....100..469M,2008MNRAS.384.1263C,2022NatAs...6..480W}. In a triple or quadruple system, the inner binary can rapidly evolve into a short-period binary through the Lidov-Kozai mechanism~\citep{1962AJ.....67..591K} combined with tidal friction~\citep{1998MNRAS.300..292K}, increasing the possibilities of a stellar merger that forms a BSS~\citep{2009ApJ...697.1048P,2014ApJ...793..137N}. However, extensive observations indicate that neither mechanism alone can fully account for BSSs detected in clusters~\citep{2009Natur.462.1028F,2023MNRAS.518L...7R}.

The rotation of BSSs is also linked to their formation and evolution. BSSs formed through binary mass transfer would gain angular momentum from their companions, leading to rapid rotation characteristics~\citep{1981A&A...102...17P,2017A&A...606A.137M}. However, the BSSs in the short-period binary systems can be spun down by tidal friction. For example, the CO-depleted BSSs in the globular cluster 47 Tucanae are slow rotators~\citep{2006ApJ...647L..53F}. Some simulations suggested that strong magnetic fields can be generated after the binary merger~\citep{2019Natur.574..211S,2022NatAs...6..480W}, which might brake the rotation of the newly formed BSSs in the subsequent evolution. The rotation velocities of BSS resulting from direct stellar collision are related to the relative velocities between the two progenitors. Head-on collision tends to produce slower BSSs, whereas off-axis collision contributes additional angular momentum to the coalescence stars, resulting in faster rotators~\citep{1997ApJ...487..290S,2005MNRAS.358..716S}. Overall, both binary evolution and stellar collision can produce BSSs that exhibit a wide range of rotation velocities, from slow to fast. Hence, it is difficult to distinguish their formation mechanisms purely by measuring their rotation. 

The formation processes of BSSs are not yet fully understood, particularly their dependence on the environment. Open clusters are ideal laboratories for stellar formation and evolution, which can provide crucial constraints on the formation of BSSs. Before the {\it Gaia} era, reliable cluster member selection was highly challenging due to the limited astrometric precision, which made the identification and study of BSSs in OCs difficult. Based on the open cluster database of  WEBDA\footnote{\url{https://webda.physics.muni.cz/}}, \citet{2007A&A...463..789A} built up a catalog of BSS candidates in Galactic OCs\footnote{Their catalog is available at \url{https://cdsarc.cds.unistra.fr/viz-bin/qcat?J/A+A/463/789}}, including 1887 BSS candidates in 427 OCs. However, only 200 stragglers (10.6\%) were classified as "1" with good membership probabilities. 

The {\it Gaia} mission \citep{GaiaCollaboration2016} provides unprecedented high-precision five astrometric parameters ($\alpha$, $\delta$, $\varpi$, $\mu_{\alpha}^*$, $\mu_{\delta}$) and three-band photometry ($G$, $G_{\rm BP}$ and $G_{\rm RP}$) \citep{2018A&A...616A...1G,2021A&A...649A...1G,2023A&A...674A...1G}, greatly improving the reliability of stellar membership determination and characterization of a large sample of OCs (for a recent review see \citealt{Cantat-Gaudin2024}). Meanwhile, with the popularity of machine learning, such as unsupervised photometric membership assignment in stellar clusters \citep[UPMASK,][]{2014A&A...561A..57K}, Density-Based Spatial Clustering of Applications with Noise \citep[DBSCAN,][]{1996A}, Hierarchical Density-Based Spatial Clustering of Applications with Noise \citep[HDBSCAN,][]{2013Density,mcinnes_hdbscan_2017}, Gaussian Mixture Models \citep[GMMs,][]{1894RSPTA.185...71P}, Friends of Friends Algorithm \citep[FOF,][]{1982ApJ...257..423H}, etc, cluster census efficiency has been tremendously raised. An increasing number of OCs have been detected based on {\it Gaia} data~\citep{Cantat-Gaudin2018,2020A&A...640A...1C,2019ApJS..245...32L,2020A&A...635A..45C,2022A&A...661A.118C,2021RAA....21...45Q,2023ApJS..265...12Q,2022ApJS..260....8H,2021A&A...646A.104H,2023A&A...673A.114H}, providing a valuable opportunity to identify BSSs in OCs and investigate their properties and formation. 

Utilizing the member stars with a membership probability greater than 0.5 from the OC catalog of \citet{2020A&A...633A..99C} based on {\it Gaia} Data Release 2 (DR2), \citet[herafter \citetalias{2021A&A...650A..67R}]{2021A&A...650A..67R} searched for BSSs among 408 OCs and finally found 897 BSS and 77 yellow straggler star (YSS) candidates\footnote{Their catalog is available in \url{https://cdsarc.cds.unistra.fr/viz-bin/cat/J/A+A/650/A67}} in 111 OCs. Additionally, they removed 39 OCs listed in the \citet{2007A&A...463..789A} catalog. Subsequently, \citet[herafter \citetalias{2021MNRAS.507.1699J}]{2021MNRAS.507.1699J} identified 868 BSSs in 228 clusters and 500 probable blue stragglers (pBSs) in 208 clusters\footnote{Their catalog is available in \url{https://cdsarc.cds.unistra.fr/viz-bin/cat/J/MNRAS/507/1699}} with the OC catalog of~\citet{2020A&A...640A...1C} based on {\it Gaia} DR2. 
In our recent work (\citealt{2023A&A...672A..81L}, hereafter \citetalias{2023A&A...672A..81L}), we focused on 541 OCs from the catalog of \citet{2022ApJS..260....8H} and implemented a uniform membership determination and isochrone fitting with {\it Gaia} Data Release 3 (DR3). Ultimately, we found 138 new BSSs in 50 OCs\footnote{Their catalog is available at \url{https://cdsarc.cds.unistra.fr/viz-bin/cat/J/A+A/672/A81}}. \citet{2024A&A...686A.215L} discovered a previously unreported BSS with reliable membership probability in the tidal tail of the intermediate age open cluster NGC~752. Based on its fast rotational velocity ($v\ \mathrm{sin}i=206.9\pm4.9$~km $\rm s^{-1}$), lack of ultraviolet excess in the spectral energy distribution, and no significant variations in the light curve, they suggested that this BSS is likely a single star formed through a stellar merger. \citet{2025arXiv250218780C} recently discovered 119 BSSs and 328 pBSs within 53 distant OCs ($R_{\rm gc} > 12$~kpc) through visual isochrone-fitting and inspection in the CMD with {\it Gaia} DR3. Until now, a total of over 2000 BSSs or pBSs have been identified in 442 OCs based on {\it Gaia} data.

\citet{2021A&A...646A.104H} have made detailed comparisons for three main algorithms (DBSCAN, HDBSCAN, GMMs) side-by-side, exploring their effectiveness in blind-searching of OCs with the all-sky {\it Gaia} dataset. They found that HDBSCAN is the most sensitive and effective algorithm for recovering open clusters in {\it Gaia} data. Afterward, with this clustering algorithm, \citet[hereafter \citetalias{2023A&A...673A.114H}]{2023A&A...673A.114H} conducted the most comprehensive OC search based on {\it Gaia} DR3 and identified 7200 cluster candidates. 749 of 4114 highly reliable clusters are newly discovered. This homogeneous OC catalog enables us to search for more BSSs in OCs with {\it Gaia} DR3 and produce the most complete BS catalog up to now. 

In this paper, we use the member catalog from \citetalias{2023A&A...673A.114H}, plot the CMD of each cluster, and visually select the OCs that might contain BSSs. We then extend the investigating area of each selected OC, reassess the membership probability with an improved {\tt UPMASK} \citep{2014A&A...561A..57K} algorithm \citep[pyUPMASK,][]{2021A&A...650A.109P}, and confirm the reliability of new BSS candidates in each OC by isochrone fitting. Eventually, we report 272 new straggler stars in 99 OCs.

The structure of this paper is organized as follows: Section~\ref{sec: search} describes the process of searching for BSSs, including visual inspection, cluster membership, isochrone-fitting, and BSS identification. In Section~\ref{sec: cat}, we present the new BSS cluster and member catalog. A discussion about the properties of newly found BSSs is given in Section~\ref{sec: dis}. Finally, we briefly summarize our results in Section~\ref{sec: sum}.

\section{Searching process}\label{sec: search}

\begin{figure*}[h!]
\centering
\includegraphics[width=0.8\textwidth, angle=0]{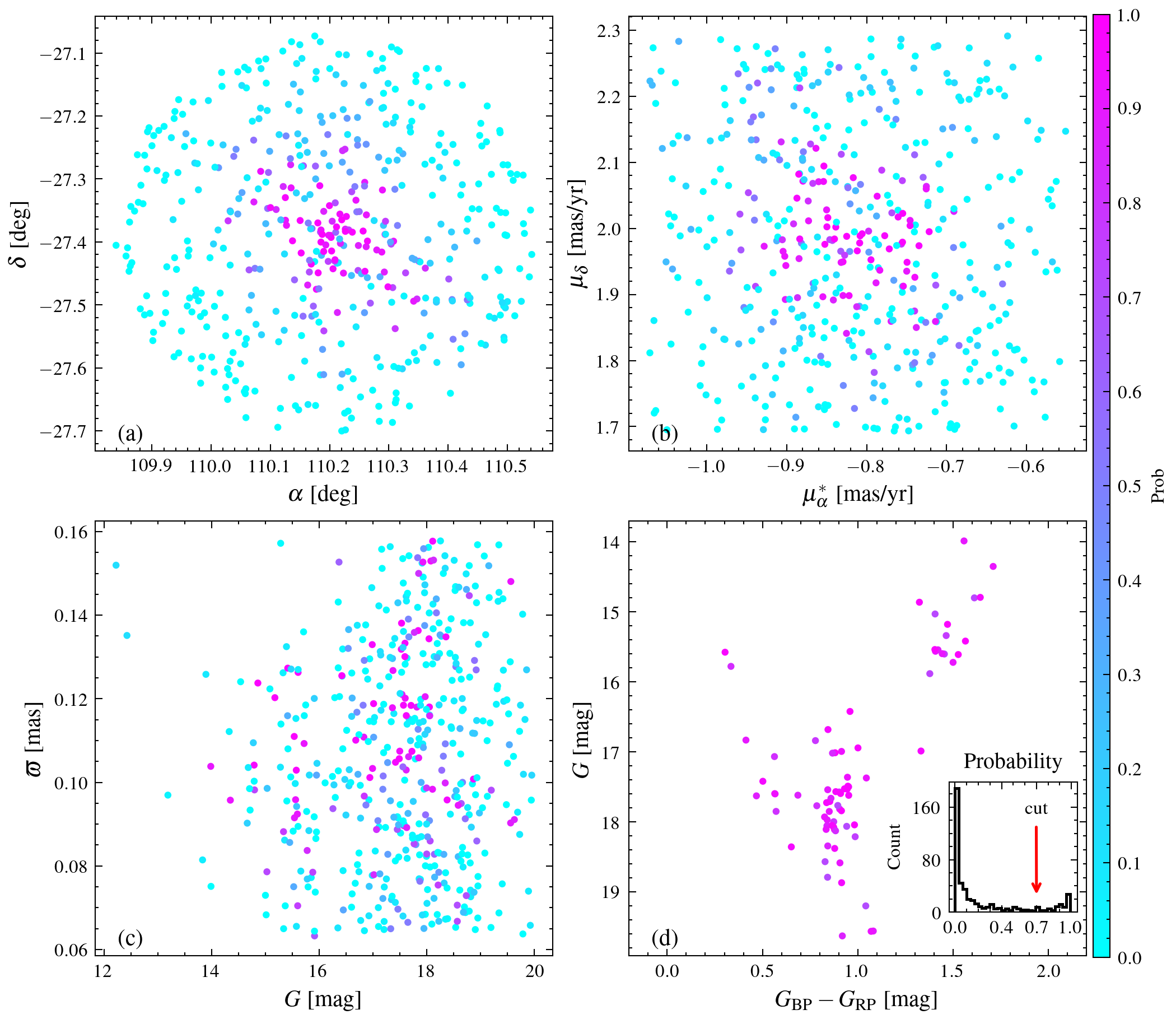}
\caption{Position (a), proper motion (b), parallax distributions (c), and color-magnitude diagram (d) of one example cluster OC\_0416. The point color represents the newly calculated membership probabilities of stars with a colorbar on the right. The subpanel inside the panel (d) is the histogram of membership probability for all stars, and the selected members in panel (d) have probabilities greater than 0.7.}
\label{fig: member}%
\end{figure*}

We search for straggler stars in OCs based on {\it Gaia} DR3, and the main steps of our work are as follows:
\begin{enumerate}
\item [1.] Initial visual inspection. We inspect the color-magnitude diagram (CMD) of each cluster in the catalog of \citetalias{2023A&A...673A.114H} and select the clusters that are likely to contain stragglers (see Section~\ref{sub: initial}).
\item [2.] Membership determination. Based on the cluster property catalog of \citetalias{2023A&A...673A.114H}, we download the data for a region around the cluster from the {\it Gaia} Archive\footnote{\url{https://gea.esac.esa.int/archive/}} and apply the {\tt pyUPMASK} algorithm to determine the membership probability for each star (see details in Section~\ref{sub: mem}).
\item [3.] Isochrone fitting. We obtain the age, distance modulus, and reddening parameters for each cluster by semi-manual isochrone-fitting with PARSEC 2.0 isochrones (Section \ref{sub: iso}).
\item [4.] Straggler identification. With the best fitting isochrone and the zero age main sequence (ZAMS), we visually inspect the CMD of each OC and confirm the BSSs, probable blue straggler stars (pBSSs), and yellow straggler stars (YSSs) -- see details in Section~\ref{sub: identify}.
\end{enumerate}

\subsection{Initial visual inspection}\label{sub: initial}

\citetalias{2023A&A...673A.114H} detected 7200 candidate clusters in {\it Gaia} DR3 and presented catalogs of cluster and member properties.
To search for stragglers in OCs, we selected 6800 clusters with type ``o'' (short for open cluster).
We downloaded the best-fitting PARSEC isochrones \citep{2022A&A...665A.126N} from CMD 3.8\footnote{\url{https://stev.oapd.inaf.it/cgi-bin/cmd}} based on their cluster property catalog for all cluster samples. Then, we visually checked the CMD of each cluster overlaid with the downloaded isochrone and then selected OCs that might contain stragglers. We crossmatched our straggler candidate cluster catalog with those from \citetalias{2021A&A...650A..67R}, \citetalias{2021MNRAS.507.1699J}, \citetalias{2023A&A...672A..81L}, and \citet{2025arXiv250218780C} and got more than 100 OCs with straggler candidates that have not been reported before.

\subsection{Membership determination}\label{sub: mem}

\begin{figure*}[h]
\centering
\includegraphics[width=\textwidth, angle=0]{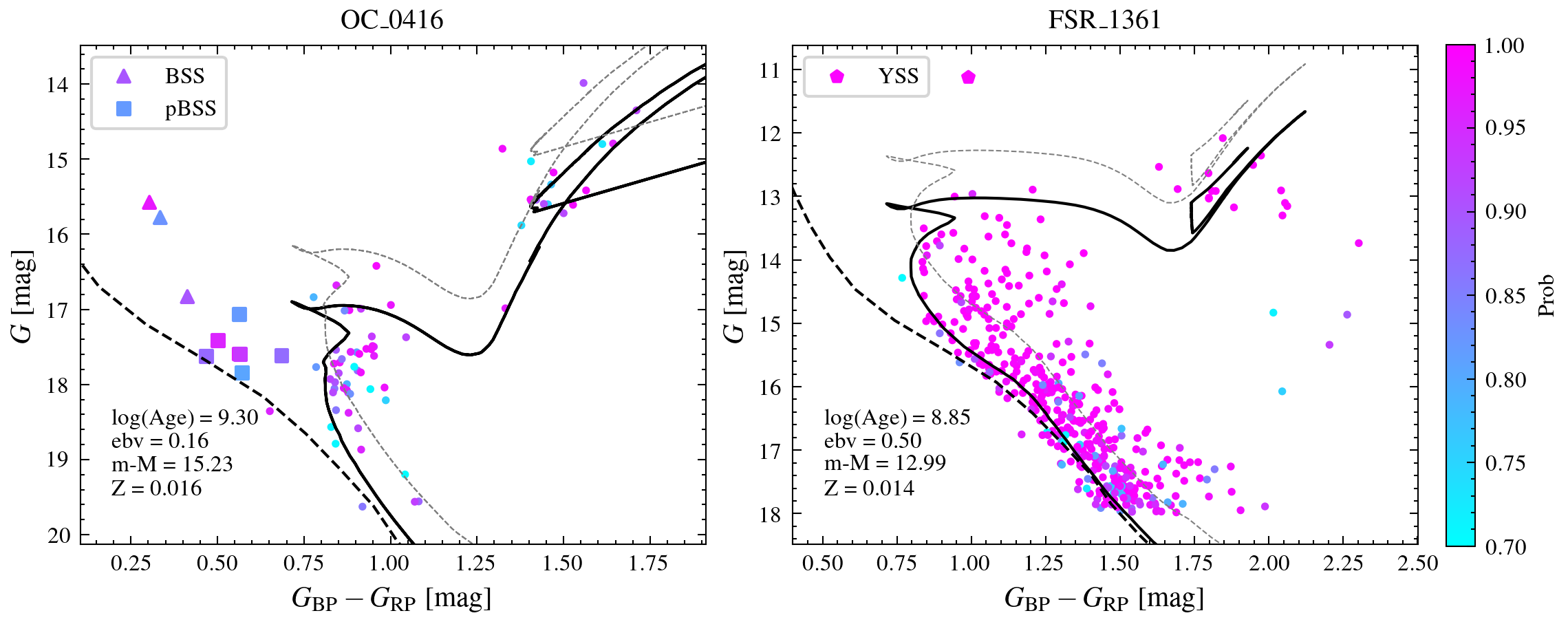}
\caption{The color-magnitude diagrams of two example open clusters OC\_0416 and FSR\_1361. The color-coded symbols are the members with membership probabilities greater than 0.7. The solid and dashed black lines represent the best-fitting isochrone and zero age main sequence (ZAMS), and the grey dashed line refers to the equal-mass binary sequence. The best-fitting age parameters are shown in the panel. The triangles, squares, and pentagon are the blue straggler stars (BSSs), probable blue straggler stars (pBSSs), and yellow straggler star (YSS) identified in our work, respectively.}
\label{fig: bs_example}%
\end{figure*}

\citet{2024A&A...686A.215L} extended the data download area for NGC\_752, and discovered an unreported BSS in the tidal tail of this intermediate age cluster. Here, to include the potential BSSs in the outer region of clusters, we uniformly expanded the data download areas to require cluster memberships for these selected clusters in Section~\ref{sub: initial}.  \citetalias{2023A&A...673A.114H} provide $\alpha$, $\delta$, $\mu_{\alpha}^{*}$, $s_{\mu_{\alpha}^{*}}$, $\mu_{\delta}$, $s_{\mu_{\delta}}$, $\varpi$, $s_{\varpi}$, $R_{\rm tot}$\footnote{($\alpha$, $\delta$) and $R_{\rm tot}$ refer to the cluster central position and total radius; ($\mu_{\alpha}^{*}$, $\mu_{\delta}$, $\varpi$) and ($s_{\mu_{\alpha}^{*}}$,  $s_{\mu_{\delta}}$, $s_{\varpi}$) refer to the mean and standard deviation values of proper motion and parallax.}, etc, in their cluster property catalog. The selection criteria of astrometric and photometric data from the {\it Gaia} Archive are as follows:
\begin{enumerate}
\item Centering on the central position ($\alpha$, $\delta$) with a radius of 50~pc if $R_{\rm tot} < 50$~pc, or $1.2 \cdot R_{\rm tot}$ if $R_{\rm tot} > 50$~pc;
\item Proper motion selection: $\mu_{\alpha}^{*} \pm 5 \cdot s_{\mu_{\alpha}^{*}}$; $\mu_{\delta} \pm 5\cdot s_{\mu_{\delta}}$;
\item Parallax selection: $\varpi \pm 3 \cdot s_{\varpi}$;
\item $G < 18$~mag form most OCs; $G < 19$~mag or 20~mag for a few distant OCs.
\end{enumerate}

After obtaining the initial sample for each cluster, we employed the {\tt pyUPMASK} algorithm \citep{2021A&A...650A.109P} to gain the membership probability of each star in the five-dimensional phase space ($l$, $b$, $\varpi$, $\mu_{\alpha}^*$, $\mu_{\delta}$). Then we visually screen each cluster in terms of position ($l$, $b$), proper motion ($\mu_{\alpha}^*$, $\mu_{\delta}$), magnitude-parallax ($G$, $\varpi$) distributions and adjusted the parameter ranges for each cluster, taking into account the membership probability cut ($P_{\rm memb} > 0.7$). Since the radius of 50~pc was too large for a few clusters, we empirically reduced it and redownloaded the initial sample. After iterating the above process twice, we ultimately adopt stars with a membership probability greater than 0.7 as the members of each cluster. For example, Fig.~\ref{fig: member} shows the property distribution of stars with different probabilities in OC\_0416, and we can clearly see the main sequence turnoff stars and giant stars of this cluster after the probability cut.

In our previous work, \citet{2022RAA....22e5022B} used ten mock samples to test the contamination and completeness of cluster members with pyUPMASK. The simulated sample includes 1200 members that are similar to the member distributions of M67 and 1200 background field stars with random and uniform distributions in the same five phase spaces. They finally derived that the average completeness of the cluster is about 97\% and the average contamination rate is 4\%. They also found that the completeness can decrease to 95\% when the $G$ magnitude reaches 18~mag. For their target cluster COIN-Gaia~13, they selected stars with membership probabilities greater than 0.5 as cluster members. In the present work, which focuses on BSSs, we adopt a more conservative membership probability threshold of 0.7 for selecting cluster members in order to minimize the field star contamination.

\subsection{Isochrone fitting}\label{sub: iso}

Our initial inspection in Section~\ref{sub: initial} of the best-fitting isochrones from \citetalias{2023A&A...673A.114H} shows their poor agreement with the observation data in the CMDs of some clusters.
To determine the ages of selected OCs, we performed new isochrone fitting with a set of non-rotation PARSEC isochrones~\citep{2022A&A...665A.126N} in the {\it Gaia} photometric system \citep{2021A&A...649A...3R} from CMD 3.8. The PARSEC isochrones have a logarithmic age grid ranging from 8.0 to 10.0 in steps of 0.05, and a metallicity grid ranging from 0.004 to 0.03 in steps of 0.002. We used the formula $A_{\rm{G}} = 2.74 \times E(B-V)$, $E(BP-RP)=1.339 \times E(B-V)$ and $E(G-RP)=0.705 \times E(B-V)$ \citep{2018MNRAS.479L.102C} to calculate the $E(B-V)$ values when conducting isochrone fitting. The fitting results of OC\_0416 and FSR\_1361 are shown in Fig.~\ref{fig: bs_example}, where the best-fitting isochrone exhibits a good match with the turnoff stars and the giant stars. The main sequence of FSR\_1361 is extended because of differential reddening.

\subsection{Straggler identification}\label{sub: identify}

Following the method of identifying BSSs in works of \citetalias{2021A&A...650A..67R}, \citetalias{2021MNRAS.507.1699J}, and \citetalias{2023A&A...672A..81L}, we checked the selected clusters individually to distinguish BSSs and YSSs. We plotted the CMD for each cluster with the best-fitting isochrone from Section~\ref{sub: iso} and the zero age main sequence (ZAMS). As shown in Fig.~\ref{fig: bs_example}, we adopted the main sequence of the best-fitting isochrones (black solid line) as the red boundary of the straggler region and the ZAMS (black dashed line) as the blue boundary.
We classified this region into blue and yellow straggler regions, based on whether the stars are bluer or redder than the turn-off point. Then we empirically divided the blue straggler region into BSS and pBSS subregions. Stars located near the main sequence turnoff point were designated as pBSSs (squares), and the remaining stars were classified as BSSs (triangles). In the yellow straggler regions, stars located above the equal-mass binary sequence, with colors between the turn-off point and the red giant branch~\citep{2004AJ....128.3019C,2021A&A...650A..67R}, were classified as YSSs (pentagon).

The spatial resolution and distance coverage of the three-dimensional dustmaps (such as {\tt Bayestar19}\footnote{\url{http://argonaut.skymaps.info/}}\citep{2019ApJ...887...93G} and {\tt stilism}\footnote{\url{http://stilism.obspm.fr}} \citep{2017A&A...606A..65C}) are limited, which prevents us from implementing reddening corrections for all cluster samples. Moreover, since the main sequences in the CMD of many clusters are not well-defined, it is not feasible to select reference stars from the lower main sequence for differential reddening correction according to the method of \citet{2012A&A...540A..16M}. Therefore, to ensure uniform identification of stragglers in OCs, we did not conduct any reddening corrections by applying dust maps or differential reddening corrections, which could potentially affect the identification of pBSSs that are closer to the turnoff region.

\begin{table*}[h]
\setlength{\tabcolsep}{7pt}
\begin{center}
\caption{\centering Description of the property catalog of OCs with stragglers.\label{tab: cat}}
\begin{tabular}{llll}
\hline\hline
Column &  Format  & Unit   & Description \\
\hline
Name & String & -- & Cluster name from Hunt23\\
$N_{\rm mem}$ & int & -- & Number of cluster members \\
$N_{\rm BSS}$& int & -- & Number of blue straggler stars\\
$N_{\rm pBSS}$ & int & -- & Number of probable blue straggler stars\\
$N_{\rm YSS}$ & int & -- & Number of yellow straggler stars \\
RA & Float & deg & Mean right ascension (J2016.0)\\
DEC & Float & deg & Mean declination (J2016.0)\\
pmRA & Float & mas~yr$^{-1}$ & Mean proper motion in right ascension direction\\
pmDEC & Float & mas~yr$^{-1}$ & Mean proper motion in declination direction\\
Plx & Float & mas & Mean parallax\\
Dist & Float & pc & Bayesian distance\\
log(Age[yr]) & Float & -- & Cluster age from isochrone fitting\\
m-M  & Float & mag  & Distance modulus from isochrone fitting\\
ebv & Float & mag & Reddening value from isochrone fitting\\
Z & Float & dex & Metallicity from isochrone fitting\\
\hline
\end{tabular}
\end{center}
\end{table*}

\begin{table*}[h]
\setlength{\tabcolsep}{7pt}
\begin{center}
\caption{\centering Description of the catalog of BSSs, pBSSs, YSSs, and all the other member stars.\label{tab: mem}}
\begin{tabular}{llll}
\hline\hline
Column &  Format  & Unit   & Description \\
\hline
source\_id & Long & -- & Unique source designation in Gaia DR3\\
ra & Float & deg & Right ascension(J2016.0)\\
dec & Float & deg & Declination (J2016.0)\\
pmra & Float & mas~yr$^{-1}$ & Proper motion in right ascension direction\\
pmra\_error & Float & mas~yr$^{-1}$ & Standard error of proper motion in right ascension direction\\
pmdec & Float & mas~yr$^{-1}$ & Proper motion in declination direction\\
pmdec\_error & Float & mas~yr$^{-1}$ & Standard error of proper motion in declination direction\\
parallax & Float & mas & Parallax \\
parallax\_error & Float & mas & Standard error of parallax \\
G & Float & mag & G band mean magnitude\\
BP & Float & mag & BP band mean magnitude\\
RP& Float & mag & RP band mean magnitude\\
Prob& Float & -- & The membership probability of stars\\
flag & String & -- & The flag indicating that the star is a possible blue straggler star (pBSS), or a bona-fide blue \\
& & &straggler star (BSS), or a yellow straggler star (YSS), or a normal member star (MEM)\\
cluster & String & -- & The corresponding cluster name\\
\hline
\end{tabular}
\end{center}
\end{table*}

\section{Results}\label{sec: cat}

In this work, we identified a total of 153 BSSs, 98 pBSSs, and 21 YSSs within 99 OCs, increasing the number of OCs with blue stragglers in the Milky Way by 22.2\% and the total number of blue stragglers by 11.2\%. We provide two catalogs\footnote{The complete catalogs are available at https://doi.org/10.57760/sciencedb.j00167.00030} with this article: one for the properties of 99 OCs with newly discovered stragglers (Table \ref{tab: cat}) and the other for parameters of their cluster members, including straggler stars (Table \ref{tab: mem}).

Based on the members with membership probabilities greater than 0.7 in Section~\ref{sub: mem}, we calculated the cluster properties, as listed in Table~\ref{tab: cat}. The first column represents the cluster name from \citetalias{2023A&A...673A.114H}. We presented the number of cluster member stars (Col. 2), BSSs (Col. 3), pBSSs (Col. 4), and YSSs (Col. 5). We also derived the mean position (Cols. 6-7), mean proper motion (Cols. 8-9), mean parallax (Col. 10) after zero-point offset correction~\citep{2021A&A...649A...4L}, and Bayesian distance\footnote{The Bayesian distance is calculated using the similar method to \citet{2025A&A...693A.317Q}: parallax zero-point offset correction~\citep{2021A&A...649A...4L} and Bayesian distance derivation~\citep{2015PASP..127..994B}} (Col. 11) for each cluster. The cluster age (Col. 12), distance modulus (Col. 13), reddening (Col. 14), and metallicity (Col. 15) were obtained from isochrone fitting in Section~\ref{sub: iso}.

Table~\ref{tab: mem} describes the catalog of cluster members, including the astrometric and photometric parameters from the {\it Gaia} DR3 (Cols. 1-12), the derived membership probabilities (Col. 13) through {\tt pyUPMASK}, the classification flag of stragglers (includes pBSS, BSS, YSS or MEM, Col. 14), and the corresponding cluster names (Col. 15).

\section{Discussion}\label{sec: dis}

In this section, we discuss the properties of the new blue stragglers (BSs: BSSs + pBSSs) we found in 99 OCs, such as the blue straggler number, frequency, density, radius, and rotational velocity.

\subsection{BS number, frequency, and density versus cluster age}

\begin{figure*}[h]
\centering
\includegraphics[width=\textwidth, angle=0]{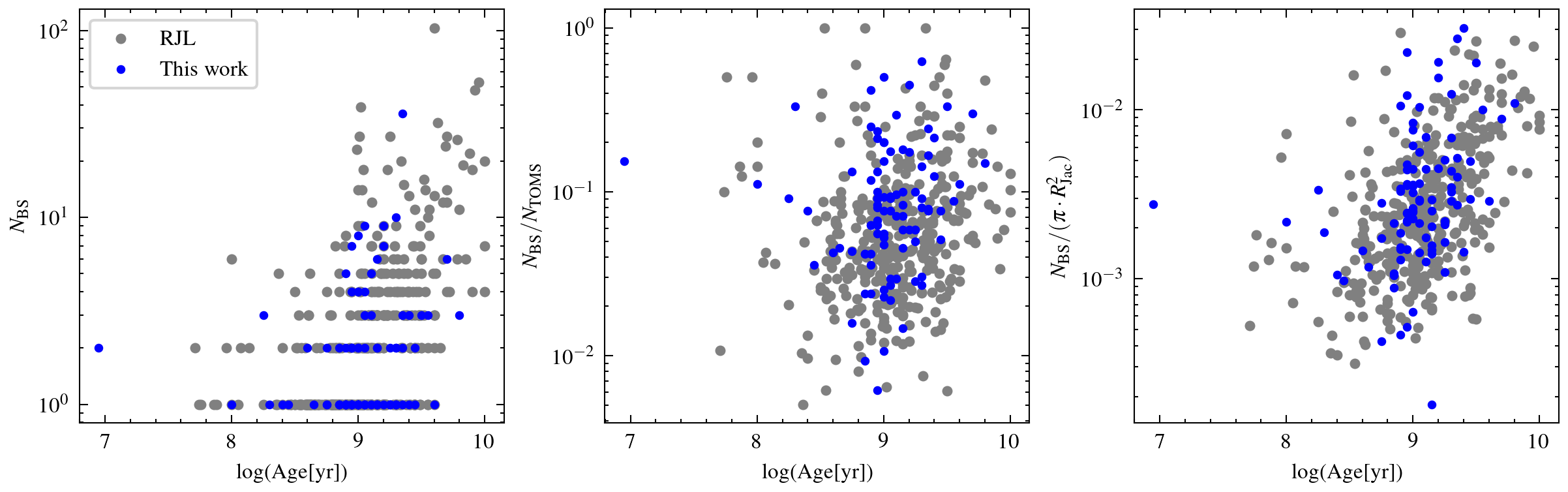}
\caption{Blue straggler number ($N_{\rm BS}$, BSS + pBSS), frequency ($N_{\rm BS} / N_{\rm TOMS}$), and density ($N_{\rm BS} / (\pi \cdot R_{\rm{Jac}}^{2})$) as a function of cluster algorithm age. The $N_{\rm TOMS}$ is the number of main sequence stars up to one magnitude below the turn-off point, and $R_{\rm{Jac}}$ is the Jacobi radius of each cluster from \citetalias{2023A&A...673A.114H}. The grey dots (RJL) refer to the sample from \citetalias{2021A&A...650A..67R} (BSS), \citetalias{2021MNRAS.507.1699J} (BSS + pBSS), and \citetalias{2023A&A...672A..81L} (BSS + pBSS). The blue dots represent the BSS + pBSS sample in this work. The cluster members used for calculating the frequencies of the RJL sample (gray dots) are from \citetalias{2023A&A...673A.114H}.}
\label{fig: N_age}
\end{figure*}

\begin{figure*}
\centering
\includegraphics[width=\textwidth, angle=0]{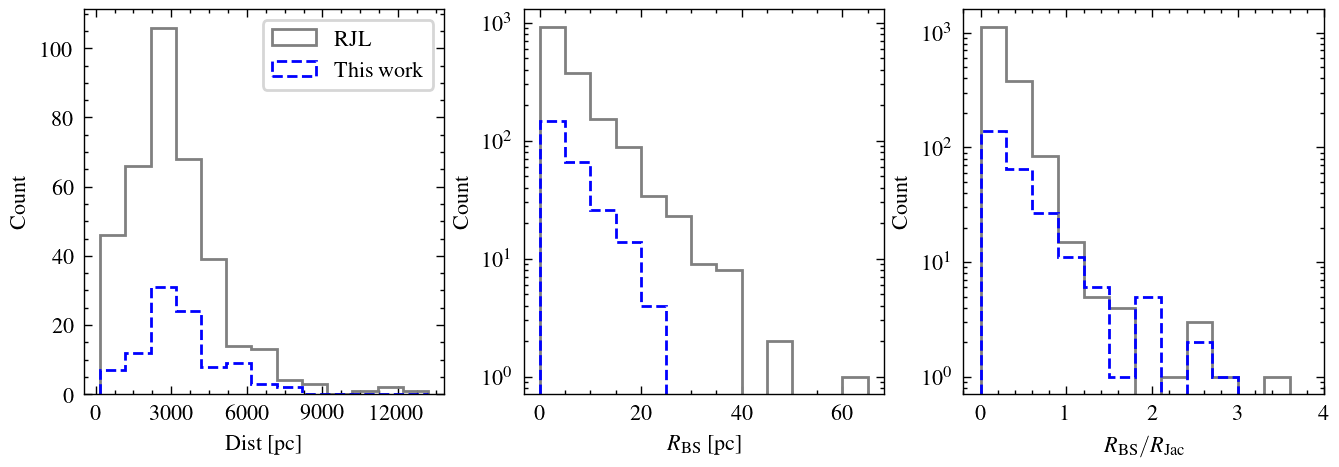}
\caption{The histograms of cluster distance (left panel), BS radial distance (middle panel), and normalized BS radial distance ($R_{\rm{BS}} / R_{\rm{Jac}}$, right panel) in OCs from \citetalias{2021A&A...650A..67R}, \citetalias{2021MNRAS.507.1699J}, and \citetalias{2023A&A...672A..81L} (RJL, grey solid line) and this work (blue dashed line). The cluster distances (Dist) of RJL sample and our sample are the `dist50' from~\citetalias{2023A&A...673A.114H} and our Bayesian distances, respectively. The BS radial distance ($R_{\rm{BS}}$) is calculated with the formula of $R_{\rm{BS}} = \rm{Dist} \times \tan{(R_{\rm{deg}})}$, where $R_{\rm{deg}}$ is the angular size from the each BS to the cluster center. $R_{\rm{Jac}}$ is the cluster Jacobi radius from \citetalias{2023A&A...673A.114H}.}
\label{fig: dist}%
\end{figure*}

Fig.~\ref{fig: N_age} shows the blue straggler number, frequency, and density as a function of cluster age, and we find similar relationships for our BS sample to those reported by \citetalias{2021A&A...650A..67R}, \citetalias{2021MNRAS.507.1699J}, and \citetalias{2023A&A...672A..81L} (RJL, grey points). As shown in the left panel, old clusters, which are typically more massive and can survive longer, tend to contain more BSs. To compare the BS frequency in different OCs, we define a normalized parameter, $N_{\rm{BS}} / N_{\rm{TOMS}}$, where $N_{\rm{BS}}$ is the number of BSs and $N_{\rm{TOMS}}$ is the number of main sequence stars up to one magnitude below the turnoff point for each OC. The middle panel illustrates the relationship between this frequency and cluster age. The BS density ($N_{\rm{BS}} / (\pi \cdot R_{\rm{Jac}}^{2})$) for each cluster is shown in the right panel, where $R_{\rm{Jac}}$ is the Jacobi radius from \citetalias{2023A&A...673A.114H}. We can see positive correlations on both panels, which indicate that older OCs have larger proportions of BSs. 

\subsection{BS radial distribution}

The first panel of Fig.~\ref{fig: dist} shows the distance distribution of OCs with BSs from \citetalias{2021A&A...650A..67R}, \citetalias{2021MNRAS.507.1699J}, and \citetalias{2023A&A...672A..81L} (RJL, grey solid line) and this work (blue dashed line). The cluster distances are adopted from the ``Dist50''\footnote{``Dist50'', maximum likelihood cluster
distances calculated using the method of \citet{2018A&A...615A..49C}} provided by \citetalias{2023A&A...673A.114H} for the RJL sample and Bayesian distances we derived for our sample in this work (see Section~\ref{sec: cat}). Similar to the reported RJL clusters, most of our cluster samples are located within a distance of 6~kpc. 

\begin{figure*}
\centering
\includegraphics[width=\textwidth, angle=0]{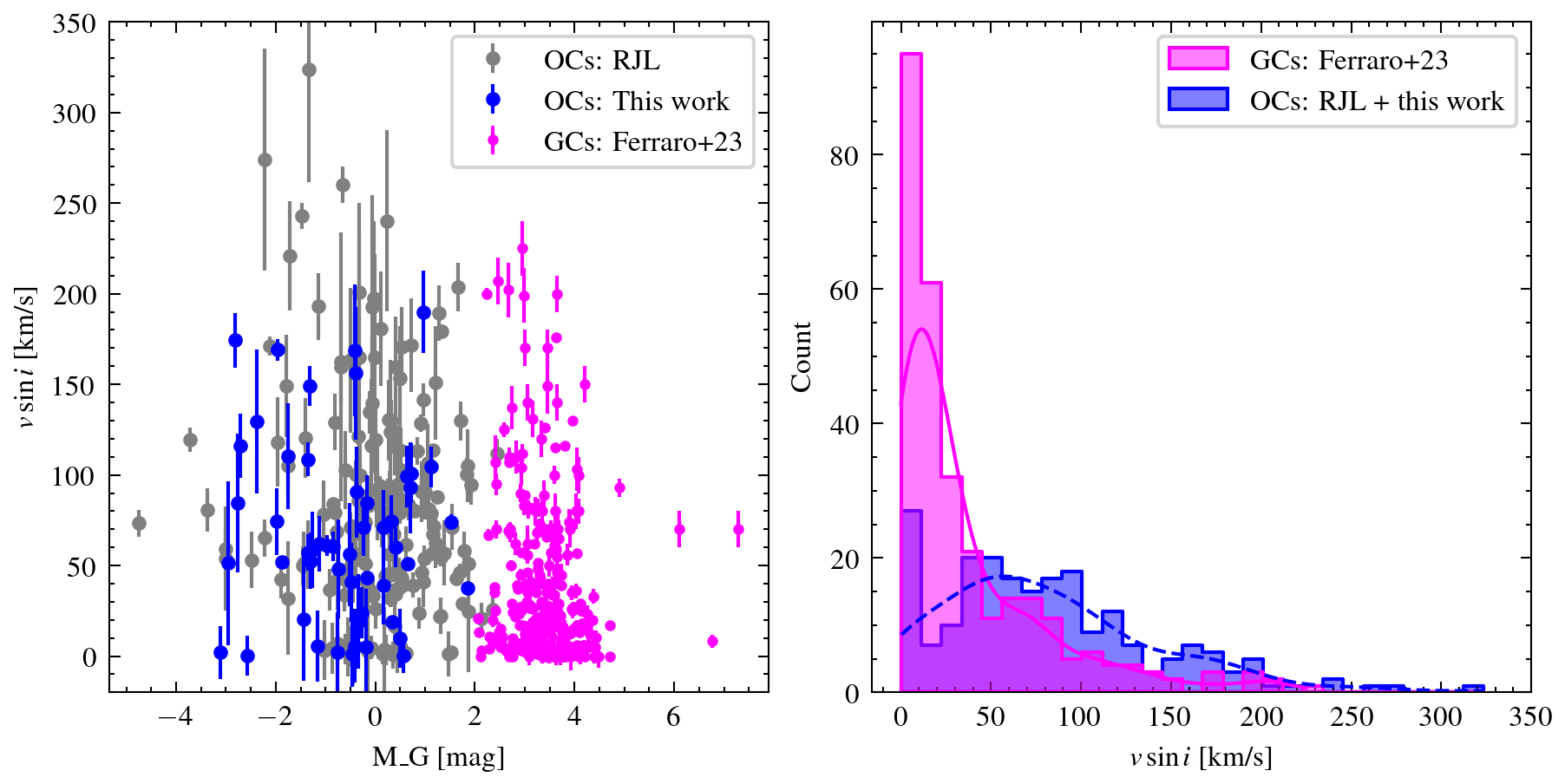}
\caption{Left panel: projected rotational velocities for BSs as a function of absolute G magnitude. Right panel: histograms of projected rotational velocities for BSs. The grey points represent the BSs in OCs of \citetalias{2021A&A...650A..67R}, \citetalias{2021MNRAS.507.1699J}, and \citetalias{2023A&A...672A..81L} (RJL), blue points are the BSs in OCs for this work, and the blue histogram represents the samples of RJL and our work. The magenta points and histogram refer to the BSs in 8 GCs from \citet{2023NatCo..14.2584F}. The blue and magenta lines refer to their corresponding kernel density distributions. The $v\sin i$ for BSs in OCs are from {\it Gaia} DR3.}
\label{fig: vsini}%
\end{figure*}

The radial distribution of BSs is influenced by both their formation channels and the dynamical evolution of their host clusters. BSs formed through stellar collisions are expected to originate in dense environments such as the GCs or the core of the most massive OCs, while those formed via binary evolution are distributed throughout the entire cluster~\citep{2024arXiv241010314W}. Dynamical processes, such as tidal fraction and two-body relaxation, can lead to mass segregation, causing more massive stars, including BSs, to migrate toward the cluster center. To investigate the radial distribution of BSs, we compute the radial distance for each BSS to the cluster center using the relation $R_{\rm{BS}} = \rm{Dist} \times \tan{(\mathit{R}_{\rm{deg}})}$, where $R_{\rm{deg}}$ is the angular separation of the BSS to the cluster center, and $\rm Dist$ is the distance plotted in the first panel. The second panel presents the histogram of the BS radial distance. Both RJL and our samples show skewed distributions, but the RJL sample includes a slightly higher fraction of BSs with radii greater than 20~pc compared to that in our work. The third panel displays the distribution of the normalized BS radial distance, $R_{\rm{BS}} / R_{\rm{Jac}}$, where $R_{\rm{Jac}}$ is the cluster Jacobi radius from \citetalias{2023A&A...673A.114H}. Both samples exhibit central peak features and then decline trends to the outer region of clusters. 

\subsection{Rotational velocities}

\citet{2023NatCo..14.2584F} measured the projected rotational velocities ($v\sin i$) of 320 BSs in 8 GCs, and found a higher fraction of fast-rotating BSs in loose clusters. We plot their determined $v\sin i$ in  Fig.~\ref{fig: vsini}, and the histogram in the right panel shows that most BSs in these 8 GCs are slow rotators with $v\sin i$ smaller than 50~km~s$^{-1}$. We crossmatch the BSs in open clusters from the RJL catalogs and this work with the astrophysical parameter catalog for hot stars from {\it Gaia} DR3 \citep{2023A&A...674A..28F}, and 215 BSs have available $v\sin i$ measurements. Their distributions are also displayed in the Fig.~\ref{fig: vsini}. In the left panel, the absolute $G$ magnitudes demonstrate that the BSs in OCs have larger masses than those in GCs. The right panel shows that the fraction of fast-rotating BSs in OCs (blue histogram) is higher than that in GCs (magenta histogram). Specifically, the fractions of fast-rotating ($ v\sin{i} > 50~\mathrm{km~s^{-1}}$) BSs in OCs and GCs are 66.5\% and 24.1\%, respectively. This discrepancy may also suggest different formation channels of BSs in these two environments: in dense GCs, BSs are more likely to be formed through binary mergers or stellar collisions, which would result in star structure reconstruction and strong magnetic fields that can lead to spin-down~\citep{2019Natur.574..211S,2022NatAs...6..480W,2024arXiv241010314W} due to mass loss and the magnetic braking effect; while in OCs, binary mass transfer plays a more important role in BS formation, producing BSs with relatively faster rotation~\citep{1981A&A...102...17P,2017A&A...606A.137M}. However, the $v\sin i$ measurements of BSSs for Galactic OCs are still incomplete. More spectroscopic observations are necessary in the future for a more detailed analysis of the $v\sin i$ distribution of BSs in these environments.

\section{Summary} \label{sec: sum}

In this study, we identified new BSSs and YSSs in 99 OCs using data from {\it Gaia} DR3, based on the cluster catalog provided by \citetalias{2023A&A...673A.114H}. Firstly, we visually inspected the CMD of each OC with the cluster members from \citetalias{2023A&A...673A.114H} to select OCs that potentially host stragglers. We then retrieved the five-dimensional astrometric and photometric data from {\it Gaia DR3} archive for these cluster candidates and acquired their membership probabilities using the {\tt pyUPMASK} package, and stars with membership probabilities greater than 0.7 were considered cluster members. Subsequently, we performed isochrone fitting for each cluster to derive the best-fit parameters, including age, distance modulus, mean reddening, and metallicity. Combining the best-fit isochrone, the ZAMS, and the equal-mass binary sequence for each OC, we identified BSSs and YSSs. In total, we discovered 153 BSSs, 98 pBSSs, and 21 YSSs within 99 OCs. This increases the number of Galactic OCs with blue stragglers by 22.2\% and the total number of blue stragglers by 11.2\%. Moreover, we analyzed the number, radial distribution, and rotational velocity properties of the newly identified BSSs. 

Similar to BSs of the RJL catalog, we found that older open clusters tend to host relatively more BSs. To compare the BS frequency in different OCs, we selected main-sequence stars up to one magnitude below the main-sequence turn-off point as reference stars. We also calculated the BS density with the cluster Jacobi radius. Both frequency ($N_{\rm BS} / N_{\rm TOMS}$) and density ($N_{\rm BS} / (\pi \cdot R_{\rm{Jac}}^{2})$) exhibit positive relationships with cluster age.

For each BS, we computed its distance to the cluster center ($R_{\rm BS}$) and found that nearly all BSs in our sample lie within a radius of 20~pc, while RJL samples included a fraction of BSs beyond 20~pc. We further calculated a normalized radial parameter ($R_{\rm BS} / R_{\rm Jac}$) using the Jacobi radius ($R_{\rm Jac}$). Both radial distance and normalized radial distance distributions exhibit central peak features and then decline trends to the outer region of clusters. 

The absolute $G$ magnitudes of most BSs in OCs are smaller than those of BSs in GCs, which is attributed to the fact that the BSs in OCs have larger masses. Using the projected rotational velocity data ($v\sin i$) from {\it Gaia} DR3, we also found that the fraction of rapid rotating BSs in OCs is higher than that of \citet{2023NatCo..14.2584F} for 8 GCs. This discrepancy may hint at the different formation mechanisms for BSs in OCs and GCs: in dense GCs, BSs are more likely to be formed through binary mergers or stellar collisions, which generate strong magnetic fields that can lead to spin-down~\citep{2019Natur.574..211S,2022NatAs...6..480W,2024arXiv241010314W} due to the magnetic braking effect; While in OCs, binary mass transfer plays a more important role in BS formation, which tends to produce BSs with relatively faster rotation~\citep{1981A&A...102...17P,2017A&A...606A.137M}.

\begin{acknowledgements}
This work is supported by the National Natural Science Foundation of China (NSFC) through grants 12090040, 12090042. Jing Zhong would like to acknowledge the science research grants from the China Manned Space Project with NO. CMS-CSST-2025-A19, the Youth Innovation Promotion Association CAS, the Science and Technology Commission of Shanghai Municipality (Grant No.22dz1202400), and Sponsored by the Program of Shanghai Academic/Technology Research Leader.
Li Chen acknowledges the science research grants from the China Manned Space Project with NO. CMS-CSST-2021-A08. This work was partially funded by the Spanish MICIN/AEI/10.13039/501100011033 and by the ``ERDF A way of making Europe'' funds by the European Union through grant PID2021-122842OB-C21, and the Institute of Cosmos Sciences University of Barcelona (ICCUB, Unidad de Excelencia `Mar\'{\i}a de Maeztu') through grant CEX2019-000918-M. FA acknowledges financial support from MCIN/AEI/10.13039/501100011033 and European Union NextGenerationEU/PRTR through grant RYC2021-031638-I.
Songmei Qin acknowledges the financial support provided by the China Scholarship Council program (Grant No. 202304910547). Chunyan Li would like to acknowledge the Natural Science Foundation Youth Program of Sichuan Province (grant No. 2025ZNSFSC0879).

This work has made use of data from the European Space Agency (ESA) mission {\it Gaia} (\url{https://www.cosmos.esa.int/gaia}), processed by the {\it Gaia} Data Processing and Analysis Consortium (DPAC, \url{https://www.cosmos.esa.int/web/gaia/dpac/consortium}). Funding for the DPAC has been provided by national institutions, in particular, the institutions participating in the {\it Gaia} Multilateral Agreement.
\end{acknowledgements}

\bibliographystyle{raa}
\bibliography{ms2025-0444}

\begin{appendix}
\section{Isochrone fitting and straggler identify results}{\label{Appendix}}

\begin{figure*}[h]
\centering
\includegraphics[width=0.75\textwidth, angle=0]{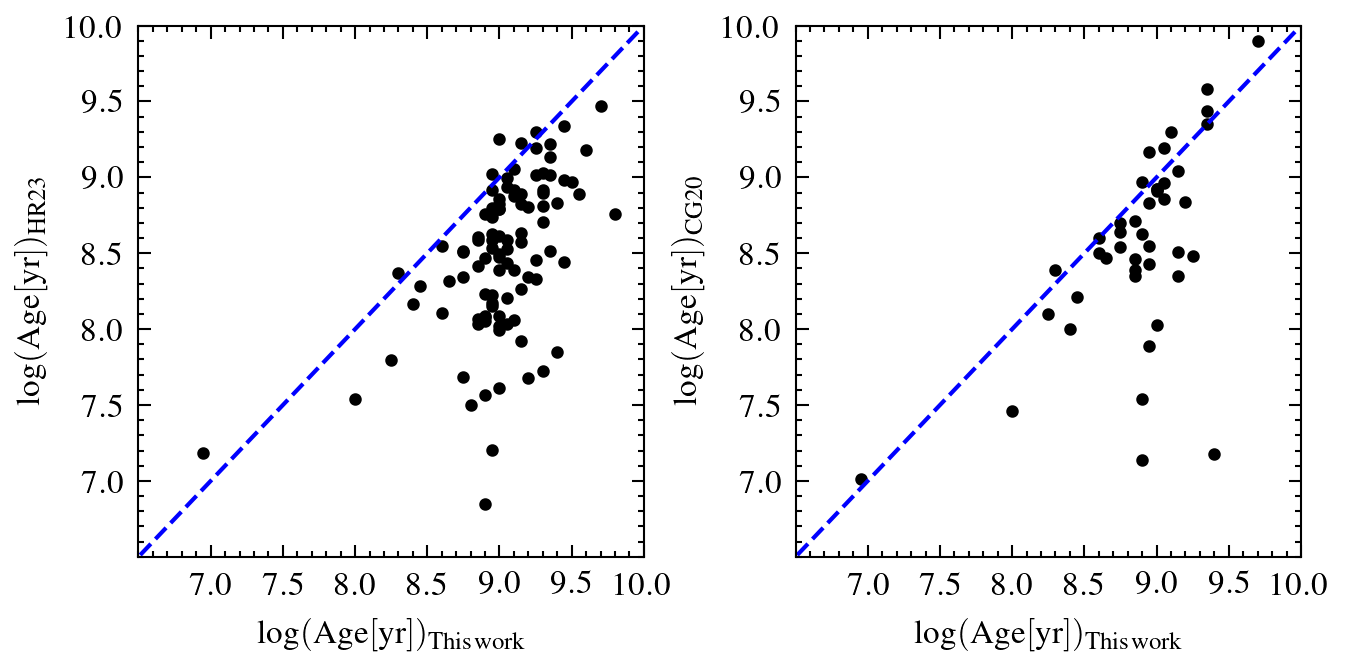}
\caption{The logarithmic age comparisons between this work and \citetalias{2023A&A...673A.114H}, \citet{2020A&A...640A...1C} (CG20), respectively.}
\label{fig: age}%
\end{figure*}

\begin{figure*}[h]
\centering
\includegraphics[width=0.9\textwidth, height=0.46\textheight, angle=0]{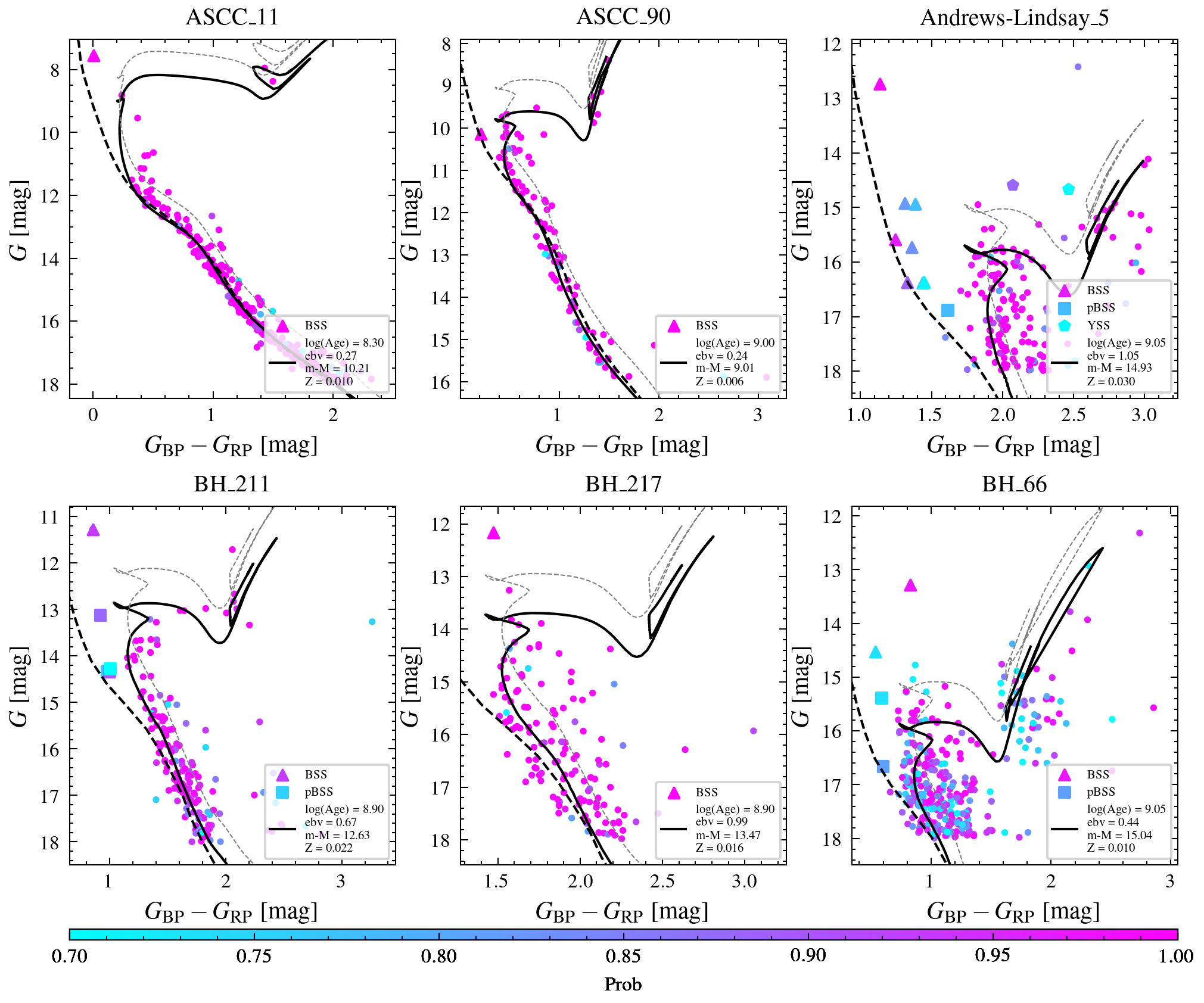}
\caption{Same as Fig.~\ref{fig: bs_example}.}
\label{fig: cmd1}%
\end{figure*}

\begin{figure*}
\centering
\includegraphics[width=\textwidth, height=\textheight, angle=0]{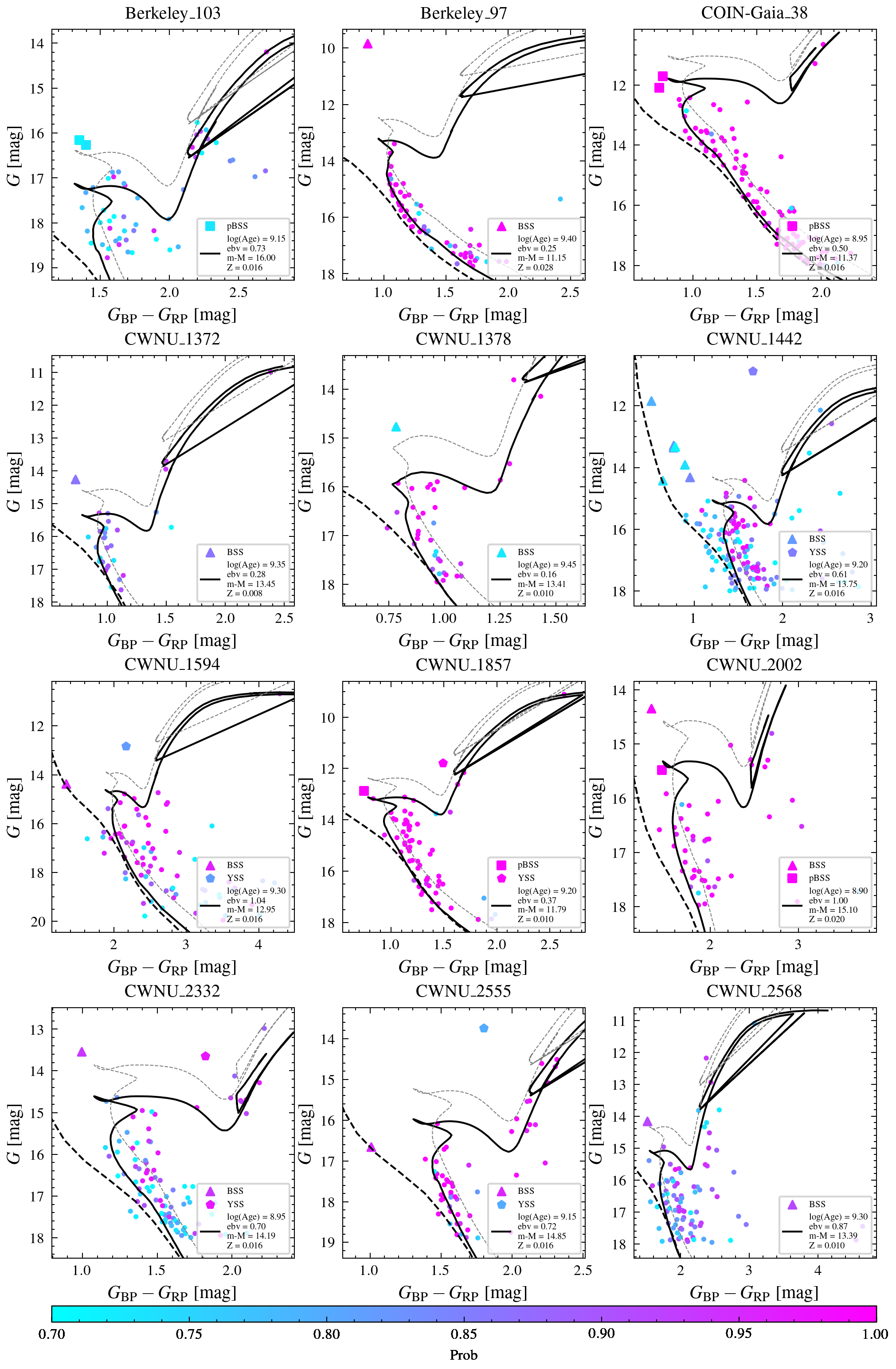}
\caption{Same as Fig.~\ref{fig: bs_example}.}
\label{fig: cmd2}%
\end{figure*}

\begin{figure*}
\centering
\includegraphics[width=\textwidth, height=\textheight, angle=0]{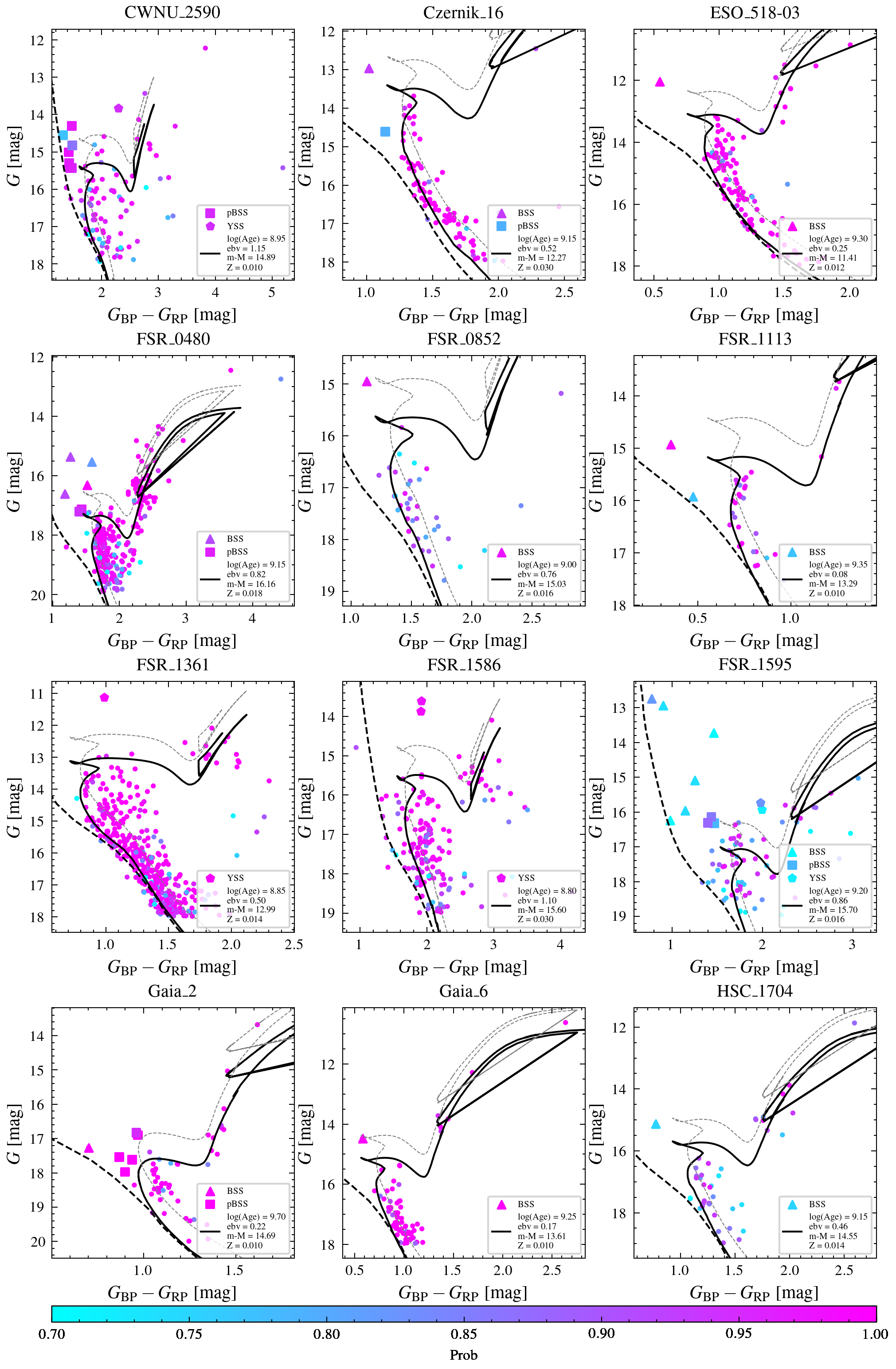}
\caption{Same as Fig.~\ref{fig: bs_example}.}
\label{fig: cmd3}%
\end{figure*}

\begin{figure*}
\centering
\includegraphics[width=\textwidth, height=\textheight, angle=0]{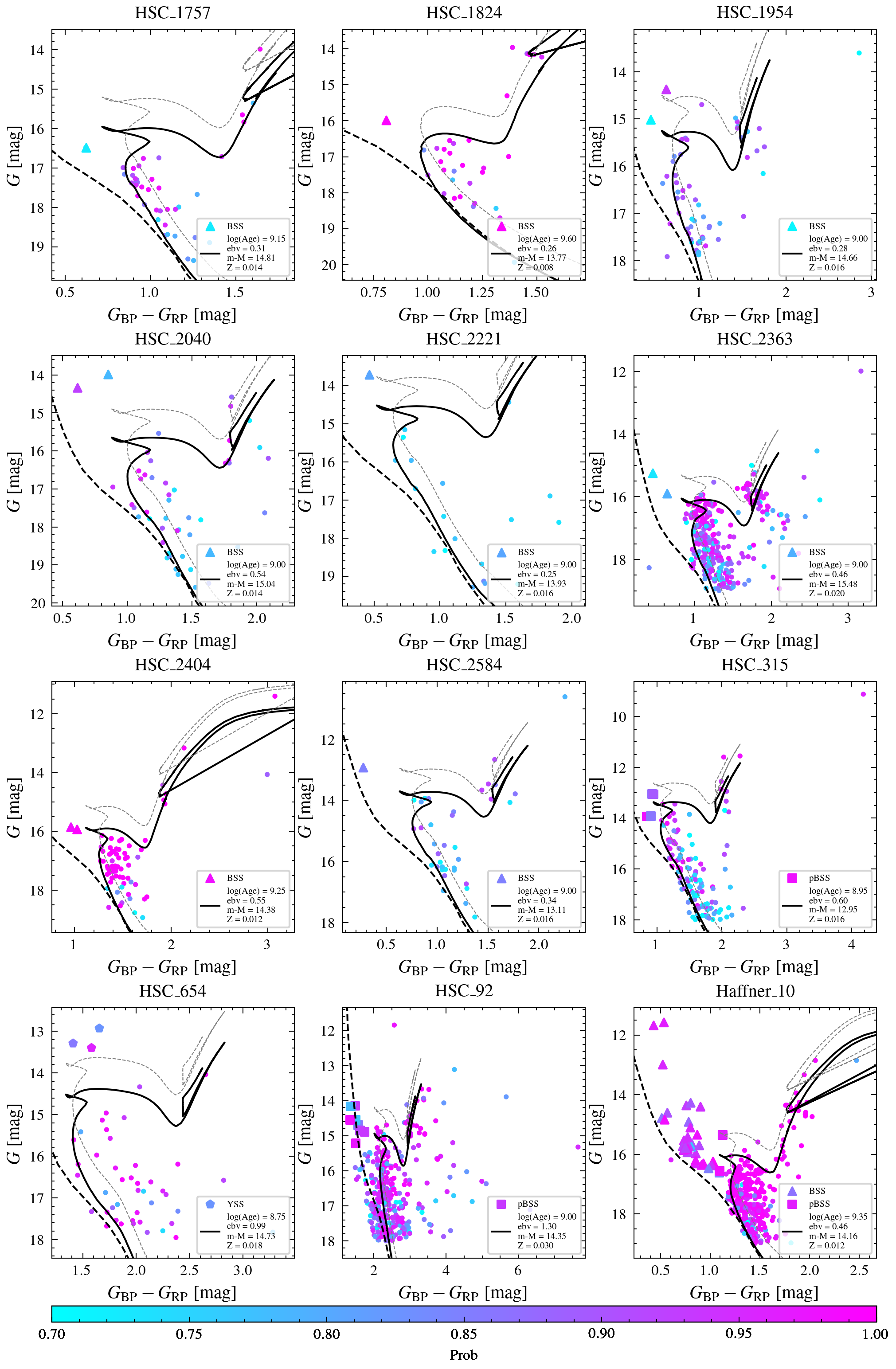}
\caption{Same as Fig.~\ref{fig: bs_example}.}
\label{fig: cmd4}%
\end{figure*}

\begin{figure*}
\centering
\includegraphics[width=\textwidth, height=\textheight, angle=0]{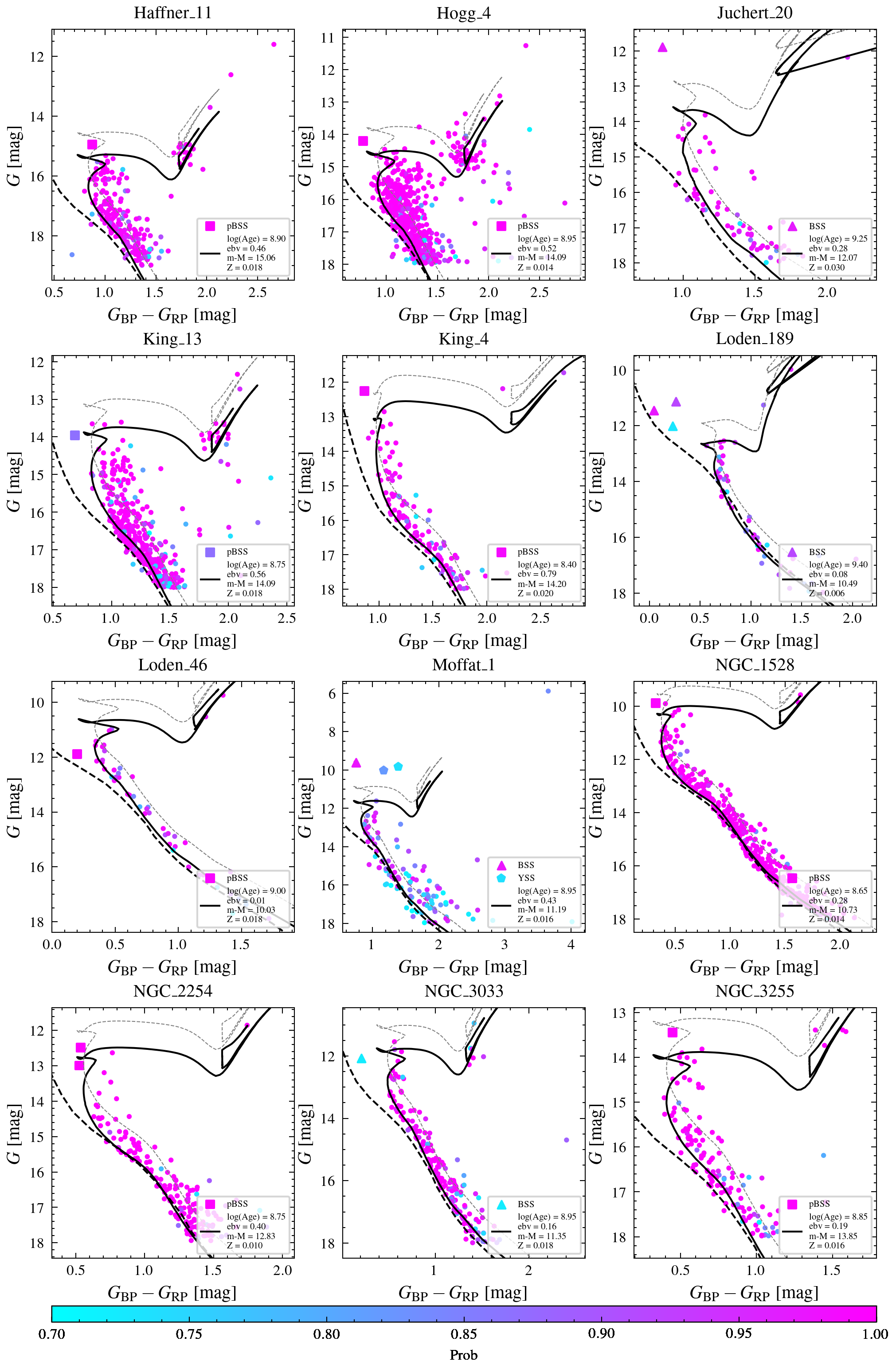}
\caption{Same as Fig.~\ref{fig: bs_example}.}
\label{fig: cmd5}%
\end{figure*}

\begin{figure*}
\centering
\includegraphics[width=\textwidth, height=\textheight, angle=0]{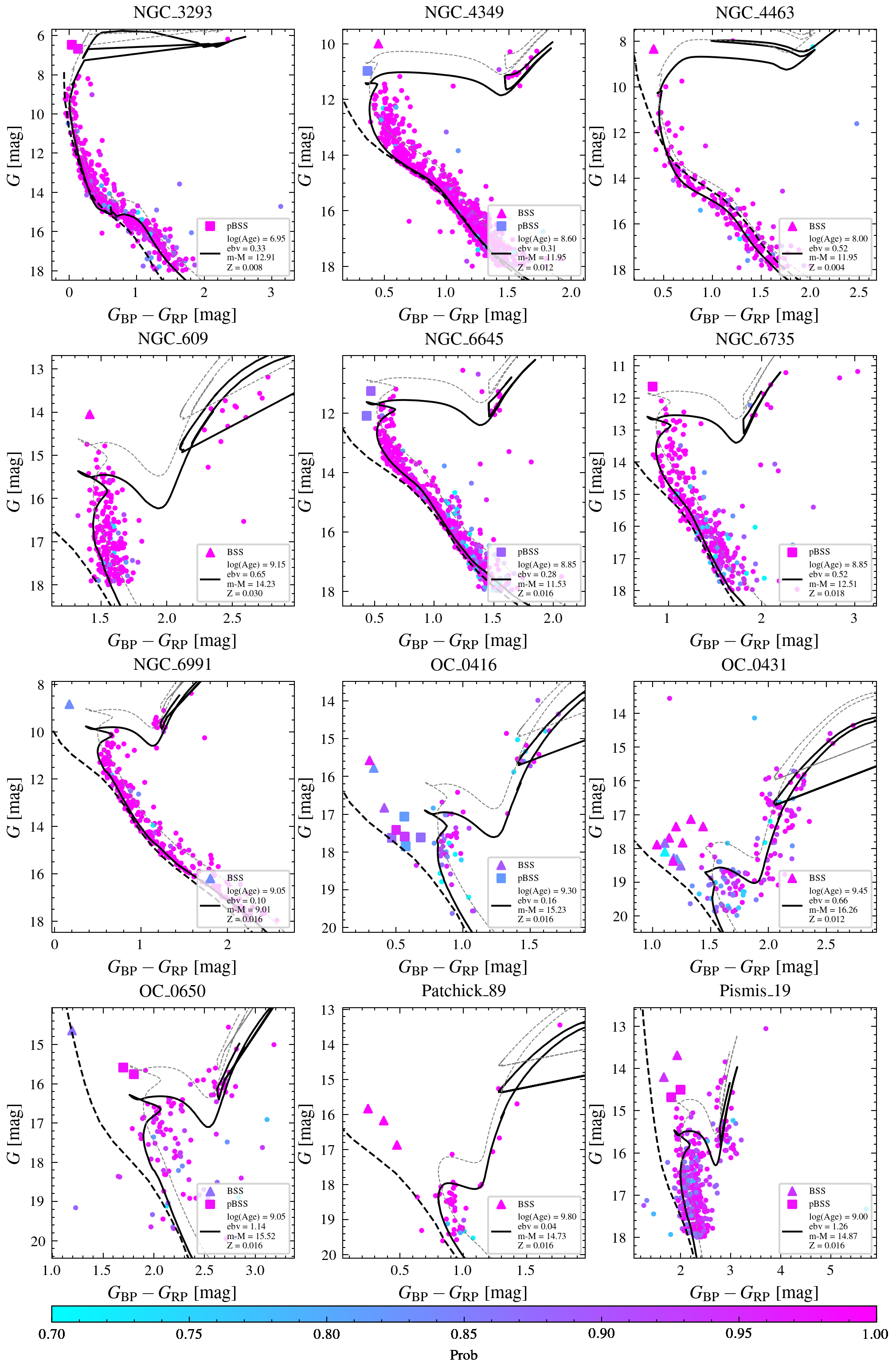}
\caption{Same as Fig.~\ref{fig: bs_example}.}
\label{fig: cmd6}%
\end{figure*}

\begin{figure*}
\centering
\includegraphics[width=\textwidth, height=\textheight, angle=0]{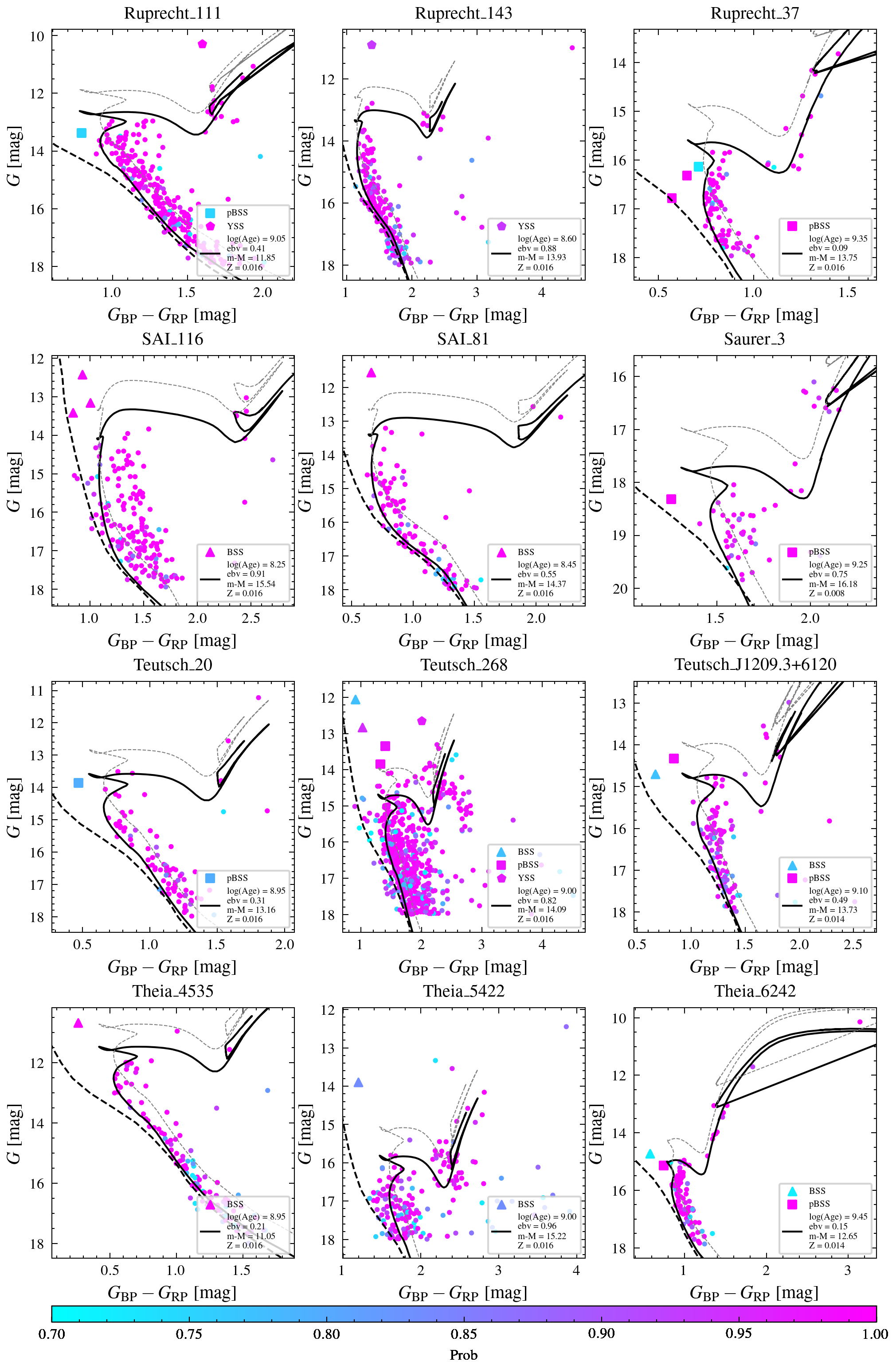}
\caption{Same as Fig.~\ref{fig: bs_example}.}
\label{fig: cmd7}%
\end{figure*}

\begin{figure*}
\centering
\includegraphics[width=\textwidth, height=\textheight, angle=0]{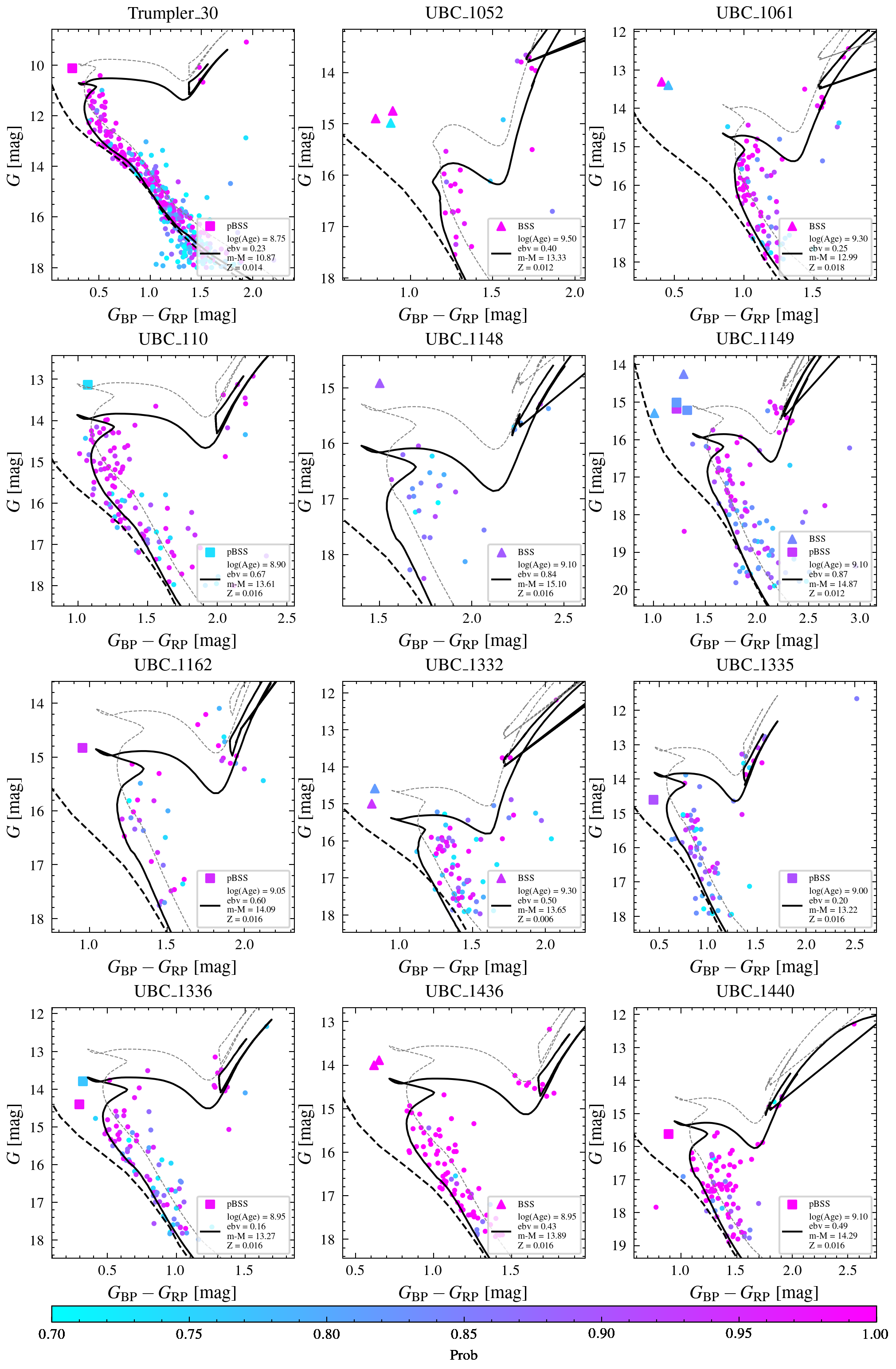}
\caption{Same as Fig.~\ref{fig: bs_example}.}
\label{fig: cmd8}%
\end{figure*}

\begin{figure*}
\centering
\includegraphics[width=\textwidth, height=0.8\textheight, angle=0]{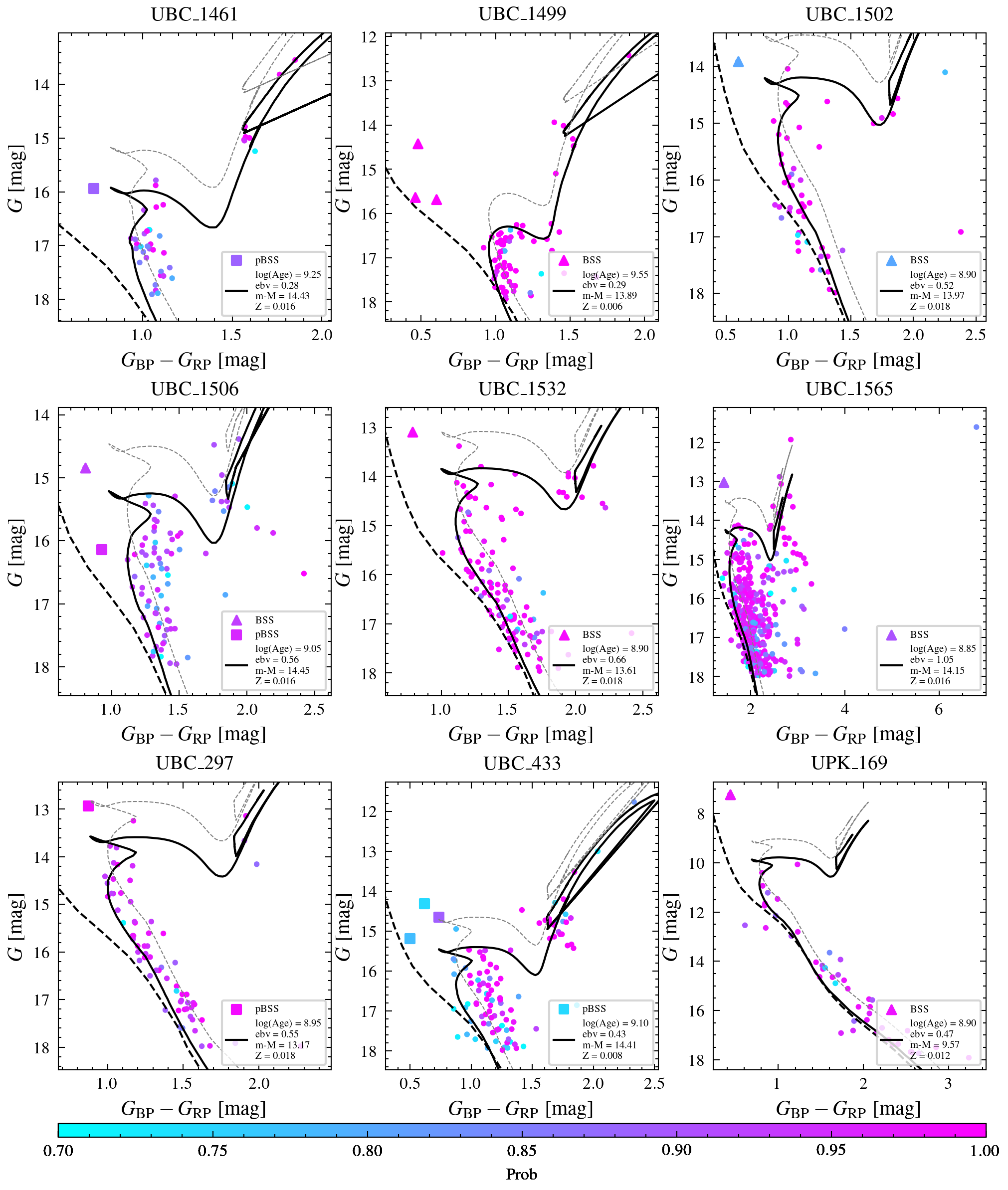}
\caption{Same as Fig.~\ref{fig: bs_example}.}
\label{fig: cmd9}%
\end{figure*}

We present the logarithmic age comparisons between this work and \citetalias{2023A&A...673A.114H}, \citet{2020A&A...640A...1C} (CG20) in Fig.~\ref{fig: age}. The age parameters we obtained are larger than those of the two previous works. We also provide the isochrone-fitting and straggler identification results of the 99 OCs in this work, as shown in Fig.~\ref{fig: cmd1} to Fig.~\ref{fig: cmd9}. The fitted isochrones show a good agreement with the observational data points.

\end{appendix}

\end{CJK}
\end{document}